\documentclass[
onecolumn,
preprintnumbers,amsmath,amssymb,prl
]{revtex4}

\usepackage{graphicx}
\usepackage{dcolumn}
\usepackage{bm}
\usepackage{amssymb}
\usepackage{epstopdf}
\usepackage{color}
\usepackage{amsmath}
\usepackage{ulem}

\begin{document}

\title{Optically induced resonant tunneling of electrons in nanostructures}
\author{M. V. Boev}
\author{V. M. Kovalev}
\author{O. V. Kibis}\email{oleg.kibis(с)nstu.ru}

\affiliation{Department of Applied and Theoretical Physics,
Novosibirsk State Technical University, Karl Marx Avenue 20,
Novosibirsk 630073, Russia}

\begin{abstract}
We developed the theory of elastic electron tunneling through a potential barrier driven by a strong high-frequency electromagnetic field. It is demonstrated that the driven barrier can be considered as a stationary  two-barrier potential which contains the quasi-stationary electron states confined between these two barriers. When the energy of an incident electron coincides with the energy of the quasi-stationary state, the driven barrier becomes fully transparent for the electron (the resonant tunneling). The developed theory is applied to describe electron transport through a quantum point contact irradiated by an electromagnetic wave.
\end{abstract}
\pacs{}

\maketitle

\section*{Introduction}
Controlling quantum systems by an off-resonant high-frequency electromagnetic field, which is based often on the Floquet theory (``Floquet engineering''), has become an established research area of modern physics~\cite{Oka_2019,Basov_2017,Goldman_2014,Bukov_2015,Casas_2001,Kibis_2020_1,Kibis_2022}. Since the off-resonant field cannot be absorbed by electrons, it only ``dresses'' them (dressing field), producing the composite electron-field states with unusual physical properties. Particularly, it has been demonstrated that the field can crucially modify electronic characteristics of various condensed-matter nanostructures, including semiconductor quantum wells~\cite{Lindner_2011,Dini_2016,Kibis_2019,Kibis_2020_2,Kibis_2021_3}, quantum rings~\cite{Kibis_2011,Koshelev_2015,Kozin_2018}, quantum dots~\cite{Kibis_2009,Kryuchkyan_2017,Iorsh_2022}, topological insulators~\cite{Rechtsman_2013,Wang_2013,Torres_2014}, carbon nanotubes~\cite{Kibis_2021_1},
graphene and related two-dimensional materials~\cite{Oka_2009,Kibis_2010,Syzranov_2013,Usaj_2014,Perez_2014,Sie_2015,Iurov_2019,Cavalleri_2020}, etc. Among a great many of nanostructures, the  nanostructures exploiting the tunneling of electrons through potential barriers --- particularly, quantum point contacts (QPCs) --- take deserved place~\cite{Imry_book,Datta_book,Wees_1988,Wees_1991,Houten_1996,Ando_1991,Rossler_2011,Thierschmann_2018,Ono_2021,Tkachenko_2015,Levin_2015,Otteneder_2018,Tkachenko_2018,Tkachenko_2021}.  Although tunnel nanostructures are one of building blocks of modern nanoelectronics, the high-frequency control of them still wait for detailed analysis. The present theoretical research is aimed to fill partially this gap. As a main result, it is found that a driven potential barrier becomes fully transparent for electrons with some energies lying below the barrier top. Such a field-induced resonant tunneling is the quantum effect which can take place in various tunnel systems, including QPCs irradiated by an electromagnetic wave (EMW).

\section*{Methods}
\begin{figure}[h!]
\centering\includegraphics[width=0.7\columnwidth]{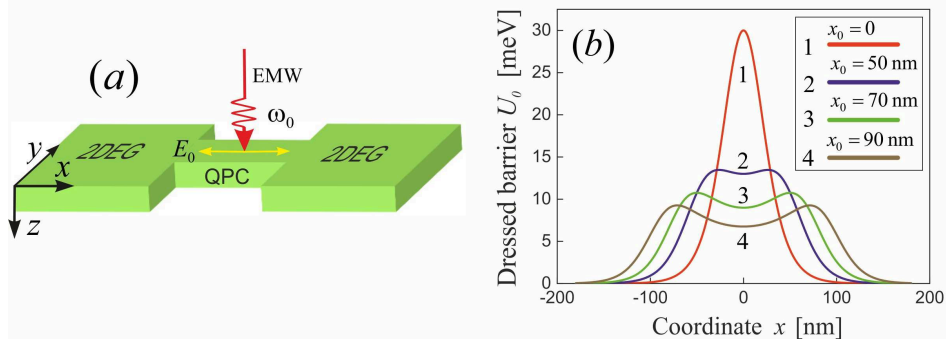}
\caption{Sketch of the system under consideration: (a) The QPC irradiated by the linearly polarized EMW with the frequency $\omega_0$ and the electric field amplitude $E_0$; (b) The dressed Eckart potential barrier with the height $u=30$~meV and the width $w=30$~nm for the different oscillation amplitudes $x_0$.}
\end{figure}
Structurally, a QPC is narrow constriction (of a width comparable to the electronic wavelength) between two wide conducting regions filled by the two-dimensional electron gas (2DEG), which is pictured schematically in Fig.~1a. In the presence of the EMW propagating along the $z$ axis and linearly polarized along the $x$ axis, the electron tunneling through the QPC can be described by the Hamiltonian
\begin{equation}\label{He}
\hat{\cal
H}_e=\frac{[\hat{{p}}_x-eA_x(t)/c]^2}{2m_e}+U(x),
\end{equation}
where $e=-|e|$ is the elementary electron charge defined as a
negative quantity, $m_e$ is the electron effective mass, $\hat{{p}}_x$ is the operator of electron momentum along the $x$ axis, and $U(x)$ is the QPC potential barrier. Assuming that the EMW length much exceeds the 2DEG thickness, the vector potential of the EMW can be written as
\begin{equation}\label{A}
A_x(t)=({cE_0}/{\omega_0})\cos\omega_0t
\end{equation}
where $E_0$ is the electric field amplitude of the EMW, and $\omega_0$ is the EMW frequency which is assumed to be far from all characteristic frequencies of the system (the off-resonant field). To optimize the tunnel problem solving, let us transform the Hamiltonian (\ref{He}) with the Kramers-Henneberger unitary transformation~\cite{Kramers_book,Henneberger_1968,Holthaus_1992}
\begin{equation}\label{K}
\hat{\mathcal{\cal U}}(t)=\exp\hspace{-0.2em}{\left[\frac{ie}{\hbar m_ec}\int^t \hspace{-0.5em} dt' \left( {A}_x(t')\hat{{p}}_x-\frac{eA_x^2(t')}{2c}\right)\right]}.
\end{equation}
Then the {exact} transformed Hamiltonian (\ref{He}) reads
\begin{equation}\label{Ht}
\hat{\cal H}=\hat{\cal U}^\dagger(t)\hat{\cal H}_e\hat{\cal U}(t) -i\hbar\hat{\cal U}^\dagger(t)\partial_t
\hat{\cal U}(t)=\frac{\hat{{p}}_x^2}{2m_e}+U(x+x_0\sin\omega_0t),
\end{equation}
where
\begin{equation}\label{X0}
x_0=\frac{|e|E_0}{m_e\omega_0^2}
\end{equation}
is the amplitude of free electron oscillations induced by the field (\ref{A}). Since the oscillation amplitude (\ref{X0}) depends on both the field amplitude $E_0$ and the field frequency $\omega_0$, we use it below as an effective parameter describing the strength of electron-field interaction. Comparing Eqs.~(\ref{He}) and (\ref{Ht}), one can conclude that the unitary transformation (\ref{K}) removes the coupling of the momentum operator $\hat{{p}}_x$ to the vector potential ${A}_x(t)$ in the Hamiltonian and transfers the oscillating time dependence from the kinetic energy of electron to its potential energy. {It should be noted that the unitary-transformed Hamiltonian (\ref{Ht}) is conventionally used to describe the electron dynamics in various periodically driven quantum systems, including the electron-atom interaction in intense laser fields~\cite{Gavrila_1984}.} Expanding
the oscillating potential $U(x+x_0\sin\omega_0t)$ into a Fourier series, the Hamiltonian (\ref{Ht}) can be rewritten as
\begin{equation}\label{H3}
\hat{\cal
H}=\frac{\hat{{p}}_e^2}{2m_e}+\sum_{m=-\infty}^\infty
U_m(x)e^{im\omega_0 t},
\end{equation}
where
\begin{equation}\label{Un}
U_m(x)=\frac{1}{2\pi}\int_{-\pi}^{\pi}U(x+x_0\sin\omega_0t)e^{-im\omega_0t}
d(\omega_0t)
\end{equation}
are the coefficients of the Fourier expansion. {The Hamiltonian \eqref{H3} is still exact and physically identical to the initial Hamiltonian \eqref{He}. Next, we have to make some approximations. In the following, we will assume that the field frequency is high enough to average the Hamiltonian (\ref{H3}) over the field period. In the high-frequency limit, the Schr\"odinger equation with the time-averaged Hamiltonian (\ref{H3}) provides a good approximation to the actual electron dynamics in various potentials~\cite{Holthaus_1992}.} For definiteness, let us approximate the potential barrier $U(x)$ by the Eckart potential~\cite{Eckart_1930},
\begin{equation}\label{EP}
U(x)=\frac{u}{\cosh^2\left(x/w\right)},
\end{equation}
where $u$ and $w$ are the effective height and width of the barrier, respectively. This model potential is often used to describe various quantum systems, including QPCs~\cite{Kay_2022,Skakala_2010,Tkachenko_2021}. Substituting Eq.~(\ref{EP}) into Eq.~(\ref{Un}), one can find the stationary part of the oscillating potential $U(x+x_0\sin\omega_0t)$, which reads
\begin{equation}\label{U0}
U_0(x)=\frac{u}{2\pi}\int_{-\pi}^{\pi}\frac{d(\omega_0t)}{\cosh^2\left[(x/w)+(x_0/w)\sin\omega_0t\right]}.
\end{equation}
The stationary potential (\ref{U0}) is the potential barrier modified by the field (the dressed barrier), which is responsible for the elastic electron tunneling discussed below. In the absence of irradiation ($x_0=0$), the dressed barrier (\ref{U0}) turns into the bare Eckart barrier (\ref{EP}) as expected. Performing numerical calculations of the integral in Eq.~(\ref{U0}), we arrive at the plots shown in Fig.~1b, which demonstrate the coordinate dependence of the dressed barrier (\ref{U0}) for the different oscillation amplitudes (\ref{X0}). {To analyze the plots, let us rewrite the dressed potential \eqref{U0} as a function
$U_0(x)=({1}/{2\pi})^2\int_{-\pi}^{\pi}d(\omega_0t)\int_{-\infty}^\infty{dq}\,e^{iq(x+x_0\sin\omega_0t)}U(q)$,
where $U(q)=\pi w^2u_0q/\sinh(\pi qw/2)$ is the Fourier image of the bare Eckart potential \eqref{EP}. Applying the Jacobi-Anger expansion, we arrive at the function $U_0(x)=(1/2\pi)\int_{-\infty}^\infty dq\,e^{iqx}J_0(qx_0)U(q)$ which has local minimum at $x=0$ under the condition}
\begin{equation}\label{ratio}
\int_{-\infty}^\infty \frac{J_0(zx_0/w)z^3dz}{\sinh\left(\pi z/2\right)}<0,
\end{equation}
where $J_0(x)$ is the zeroth order Bessel function of the first kind. If the ratio $x_0/w$ is large enough to satisfy the condition \eqref{ratio}, the irradiation turns the Eckart potential barrier (\ref{EP}) plotted with the curve $1$ in Fig.~1b into the two-barrier potential (see the curves $2$--$4$ in Fig.~1b corresponding to the non-zero irradiation). The two-barrier structure of the dressed potential barrier (\ref{U0}) results in the features of electron tunneling analyzed below.

\section*{Results and discussion}
The elastic electron transport through the QPC pictured schematically in Fig.~1a can be described by the well-known Landauer formulae (see., e.g., Ref.~\onlinecite{Datta_book})
\begin{equation}\label{L}
J=\frac{2e}{h}\int\limits_0^\infty d\varepsilon D(\varepsilon)[f(\varepsilon)-f(\varepsilon+eV)],
\end{equation}
where $J$ is the electric current through the QPC, $V$ is the voltage applied between the two 2DEG regions, $f(\varepsilon)$ is the Fermi-Dirac distribution function for electrons in these regions, $\varepsilon$ is the electron energy, $D(\varepsilon)={j_t}/{j_i}$ is the transmission coefficient describing the probability of electron tunneling through the potential barrier (the barrier transparency), whereas $j_i$ and $j_t$ are the probability currents corresponding to the electron waves incident to the barrier and transmitted through the barrier, respectively. These currents are defined by the conventional equation, $j=-({i\hbar}/{2m_e})\left(\psi^*{\partial_x\psi}-\psi{\partial_x \psi^*}\right)$,
where $\psi$ are the eigenfunctions of the stationary tunnel Hamiltonian,
$\hat{\cal H}_0={\hat{{p}}_e^2}/{2m_e}+U_0(x)$, which satisfy the Schr\"odinger equation $\hat{\cal H}_0\psi=\varepsilon\psi$. It follows from Eq.~(\ref{L}) that the dependence of the barrier transparency $D(\varepsilon)$ on the field (\ref{A}) should be found to describe the electron transport through the irradiated QPC.

\begin{figure}[h!]
\centering\includegraphics[width=0.4\columnwidth]{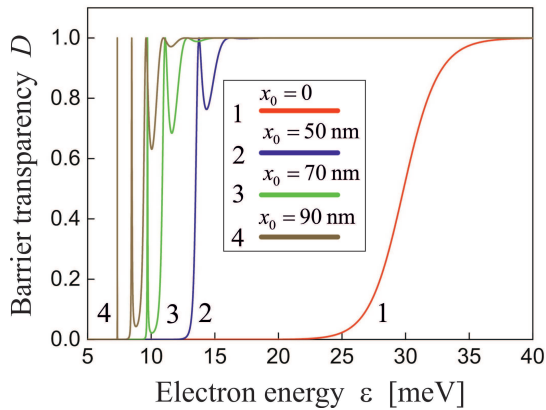}
\caption{Barrier transparency $D$ as a function of electron energy $\varepsilon$ for different oscillation amplitudes $x_0$  in the GaAs-based QPC with the electron effective mass $m_e=0.067m_0$ and the Eckart potential barrier with the height $u=30$~meV and the width $w=30$~nm.}
\end{figure}
In the absence of the irradiation, the barrier transparency is described by the known expression for the Eckart potential~\cite{Eckart_1930},
\begin{equation}\label{DE}
D(\varepsilon)=\frac{\sinh^2\left(\sqrt{{\varepsilon}/{\varepsilon_e}}\right)}{\cosh^2\left(\sqrt{{u}/{\varepsilon_e}-{\pi^2}/{4}}\right)+\sinh^2\left(\sqrt{{\varepsilon}/{\varepsilon_e}}\right)},
\end{equation}
where $\varepsilon_e=\hbar^2/2m_e\pi^2w^2$ is the characteristic energy of an electron in the Eckart potential. In order to find the transparency $D(\varepsilon)$ in the presence of irradiation, the tunnel problem with the dressed potential barrier (\ref{U0}) should be solved numerically. The found dependence of the barrier transparency $D(\varepsilon)$ on the electron energy is plotted in Fig.~2 for the different oscillation amplitudes (\ref{X0}). It follows from Eq.~(\ref{DE}) that the transparency monotonically depends on the electron energy in the absence of irradiation (see the curve $1$ in Fig.~2), whereas the irradiation results in its oscillating behaviour with $D(\varepsilon)=1$ for some energies $\varepsilon$ (see the curves $2$--$4$ in Fig.~2). Physically, these features of the barrier transparency originate from the quasi-stationary electron states confined between the two barriers of the dressed potential (\ref{U0}). When the energy of an incident electron coincides with the energy of the quasi-stationary state, the electron waves reflected from these two barriers propagate in antiphase and, correspondingly, suppress each other due to the destructive interference. As a consequence, the barrier (\ref{U0}) becomes fully transparent for the electron (the resonant tunneling~\cite{Bohm_book}). Within the semiclassical Wentzel–Kramers–Brillouin (WKB) approach, the energies of the quasi-stationary states, $\varepsilon_n$, can be found as solutions of the Bohr-Sommerfeld equation
\begin{equation}\label{En}
\frac{1}{\pi\hbar}\int_{-a}^adx\sqrt{2m_e[\varepsilon_n-{U}_0(x)]}=n+\frac{1}{2},
\end{equation}
where $n=0,1,2,...$ are the numbers of the states. Correspondingly, the broadenings of the energies are $\Gamma_n={\hbar}/{\tau_n}$, where $\tau_n={T_n}/{2D_n}$ is the lifetime of the state,
\begin{equation}\label{D}
D_n=\exp\left(-{2}\int_{a}^bdx\sqrt{2m_e[{U}_0(x)-\varepsilon_n]}/\hbar\right)
\end{equation}
is the transparency of each of the two barriers,
\begin{equation}\label{T}
T_n=\int_{-a}^a\frac{dx\sqrt{2m_e}}{\sqrt{\varepsilon_n-{U}_0(x)}}
\end{equation}
is the period of classical electron motion between these barriers, and $\pm a,\pm b$ are the coordinates of the turning points (see the insert in Fig.~3a). Near the resonances $\varepsilon=\varepsilon_n$, the dependence of the barrier transparency on electron energy acquires the Breit-Wigner form,
\begin{equation}\label{BW}
D(\varepsilon)=\frac{(\Gamma_n/2)^2}{(\varepsilon-\varepsilon_n)^2+(\Gamma_n/2)^2},
\end{equation}
which appears often in the theory of scattering through quasi-stationary states~\cite{Kibis_2020_2,Coon_1989}. It follows from Eq.~(\ref{BW}), particularly, that the barrier is fully transparent, $D(\varepsilon_n)=1$, for $\varepsilon=\varepsilon_n$. Solving Eq.~(\ref{En}) numerically, we arrive at the energy levels $\varepsilon_n$ and the energy broadenings $\Gamma_n$ plotted in Fig.~3 as a function of the oscillation amplitude (\ref{X0}). Particularly, it follows from Fig.~3 that the broadening $\Gamma_n$ increases with increasing the quasi-stationary state energy $\varepsilon_n$. As a consequence, the oscillations plotted in Fig.~2 are most pronounced for small electron energies $\varepsilon$. It should be noted that the plots in Fig.~2 correspond to exact numerical solution of the tunnel problem with the dressed potential barrier (\ref{U0}), whereas the plots in Fig.~3 were obtained within the semiclassical WKB approach which is appropriate only for the approximate analysis. Particularly, the semiclassical Bohr-Sommerfeld equation (\ref{En}) describes only the resonant energies $\varepsilon_n$ lying below the top of the dressed barrier $U_0(x)$. However, there are also the quasi-stationary states lying above the top of the dressed potential (\ref{U0}). Physically, they are confined between two barriers of the dressed potential (\ref{U0}) due to the electron reflection above the barriers, what is the purely quantum effect. Therefore, these states --- and the resonant tunneling through them --- cannot be described accurately within the semiclassical WKB approach.
\begin{figure}[h!]
\centering\includegraphics[width=1\columnwidth]{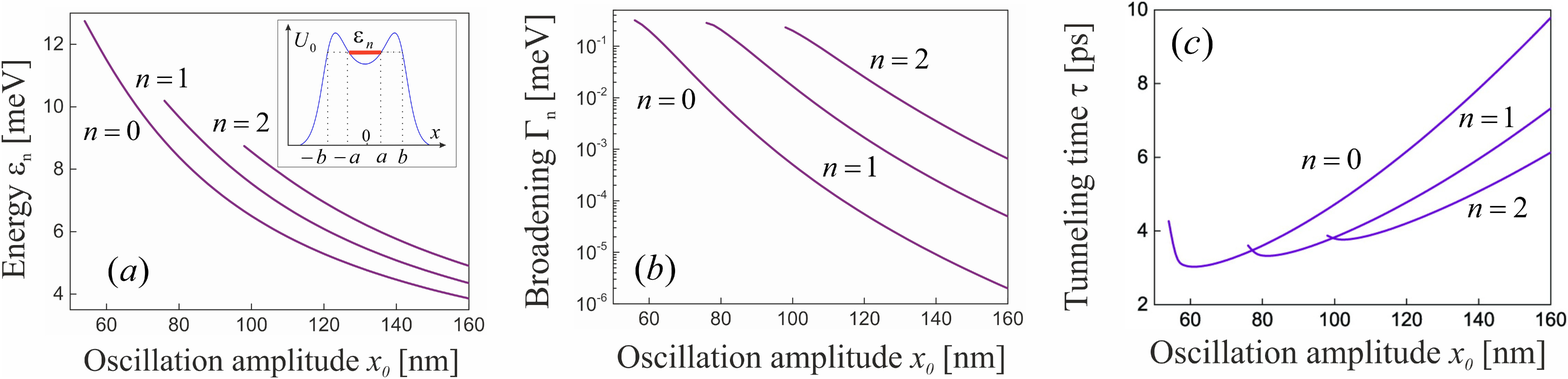}
\caption{(a) The energies of the quasi-stationary states $\varepsilon_n$, (b) their broadenings $\Gamma_n$ and (c) the time of resonant tunneling through these states as a function of the oscillation amplitude $x_0$ for the GaAs-based QPC with the electron effective mass $m_e=0.067m_0$ and the Eckart potential barrier with the height $u=30$~meV and the width $w=30$~nm. The insert shows schematically the quasi-stationary electron state with the energy $\varepsilon_n$, which is confined between the two barriers of the dressed potential $U_0(x)$.}
\end{figure}

To complete the analysis of irradiated QPCs (see Fig.~1a), let us discuss how the considered effects can be observed experimentally. In the linear regime with respect to the voltage $V$ applied to the QPC and for relatively small temperatures $T\ll \varepsilon_F$, the Landauer formulae (\ref{L}) results in the QPC volt-ampere characteristic $J=(2e^2/h)D(\varepsilon_F)V$. It follows from this that the irradiation-induced oscillations of the barrier transparency $D(\varepsilon)$ will result in the oscillating behaviour of the electric current through the QPC as a function $J(V_g)$, where $V_g$ is the gate voltage applied to the 2DEG to tune the Fermi energy $\varepsilon_F$. Since the peaks of the oscillations correspond to the resonant electron tunneling trough the QPC, the conductance of the QPC in the peaks, $G=J/V$, will be equal to the conductance quantum $G_0=2e^2/h$.

{It should be noted that the known theory of photon-assisted tunneling in nanostructures (see, e.g.,  Ref.~\onlinecite{Platero_2004} and references therein) describes the inelastic processes accompanied by the field absorption. Mathematically, they originate from the oscillating terms of the Hamiltonian (\ref{H3}) (i.e. the Fourier expansion terms (\ref{Un}) with $m\neq0$). On the contrary, the present theory describes the elastic resonant tunneling which originates from the stationary term of the total tunnel Hamiltonian (\ref{H3}) and, therefore, cannot be accompanied by the field absorption. Normally, both the elastic resonant tunneling and the inelastic photon-assisted tunneling processes coexist. Let us discuss the physical situations, where the inelastic tunneling processes are suppressed and, correspondingly, the considered resonant tunneling is dominant.} Evidently, the contribution of the oscillating terms of the Hamiltonian (\ref{H3}) to the tunneling can be neglected if the traversal time for tunneling through the potential barrier (tunneling time), $\tau$, much exceeds the field period $T=2\pi/\omega_0$. Indeed, if the period $T$
is very less as compared to the time $\tau$ during which an electron interacts with the barrier, then the electron
``feels'' only the time-averaged static barrier (\ref{U0}) during its traversal. Thus, the applicability condition of the developed theory is
\begin{equation}\label{tau1}
\omega_0\tau\gg1.
\end{equation}
Following B\"uttiker and Landauer~\cite{Buttiker_1982}, the traversal time for tunneling of an electron with the energy $\varepsilon$ through a potential barrier $U_0(x)$ reads
\begin{equation}\label{tau}
\tau=\sqrt{\frac{m_e}{2}}\int_{x_1}^{x_2}\frac{dx}{\sqrt{|U_0(x)-\varepsilon|}},
\end{equation}
where $x_{1,2}$ are the turning points satisfying the equality $U_0(x_{1,2})=\varepsilon$. Applying Eq.~(\ref{tau}) to describe the traversal time for resonant tunneling through the quasi-stationary state with the energy $\varepsilon_n$ (see the insert in Fig.~3a), we arrive at the equality
\begin{equation}\label{t}
\tau=\sqrt{\frac{m_e}{2}}\left[\int_{-b}^{-a}\frac{dx}{\sqrt{U_0(x)-\varepsilon_n}}\right.
+\left.\int_{-a}^{a}\frac{dx}{\sqrt{\varepsilon_n-U_0(x)}}+\int_{a}^{b}\frac{dx}{\sqrt{U_0(x)-\varepsilon_n}}\right],
\end{equation}
which has the clear physical meaning: The first and third terms describe the traversal time for tunneling of an electron through the two potential barriers confining the quasi-stationary state, whereas the second terms is the semiclassical travelling time of the electron between these barriers within the quasi-stationary state. The time (\ref{t}) as a function of the oscillation amplitude $x_0$ is plotted in Fig.~3c for the first three quasi-stationary states ($n=0,1,2$). Taking into account the known dependence of the oscillation amplitude (\ref{X0}) on the field amplitude $E_0$ and the field frequency $\omega_0$, these plots can be used to find the time $\tau$ --- and, correspondingly, to calculate the applicability condition (\ref{tau1}) --- for different fields. Last, let us estimate the contribution of the oscillating terms of the Hamiltonian (\ref{H3}) to the inelastic photon-assisted tunneling accompanied by absorption of photons of the driving field. In the basis of plane electron waves $\psi_k=e^{ikx}$, the matrix elements of these terms for the Eckart potential barrier (\ref{EP}) read
\begin{equation}\label{H4}
\langle\psi_{k'}|U_m(x)|\psi_k\rangle=\frac{(-1)^n{\pi}w^2uq\,J_m(qx_0)}{\sinh\left(\pi qw/2\right)},
\end{equation}
where $q=k'-k$, and $J_m(qx_0)$ is the Bessel function of the first kind. To take into account the effect of these terms on the electron transport through the irradiated QPC accurately, the Floquet-Landauer formalism~\cite{Moskalets_2002} can be applied. However, it follows from Eq.~(\ref{H4}) that the probability of the photon-assisted processes rapidly decreases with increasing the field frequency $\omega_0$. Particularly, the probability of the photon-assisted tunneling accompanied by absorption of $m$ photons, $W_m(\omega_0)$, becomes exponentially small in the high-frequency limit:
\begin{equation}\label{W}
W_m(\omega_0)|_{\omega_0\rightarrow\infty}\propto|\langle\psi_{k'}|U_m(x)|\psi_k\rangle|^2_{q=\sqrt{2m_em\omega_0/\hbar}}
\propto\exp{\left(-\sqrt{m\omega_0/\omega_e}\right)},
\end{equation}
where $\omega_e=\hbar/2m_e\pi^2w^2$ is the characteristic frequency of an electron in the potential (\ref{U0}). As a consequence, the oscillating terms of the Hamiltonian (\ref{H3}) can be omitted if the field frequency is high enough. Thus, the discussed resonant tunneling is dominant tunneling process in the high-frequency regime under consideration. It should be noted that the present theory does not take into account the processes which can break the phase coherence of the electron wave and, therefore, are crucial for the resonant tunneling. Such processes always take place in real conducting systems (e.g., the inelastic scattering of conduction electrons by phonons). Therefore, the developed theory is correct if the mean free path of a conduction electron in such processes (the length of the phase coherence of the electron wave) much exceed the width of the quantum point contact, $w$. Thus, the width $w$ in experiments  should be small enough.

It should be stressed that the Schr\"odinger problem with the stationary potential (\ref{U0}) contains the only electron-field parameter --- the amplitude of free electron oscillations (\ref{X0}) under the field (\ref{A}) --- which depends on both the field frequency $\omega_0$ and the field amplitude $E_0$. Therefore, the calculation results shown in Figs.~1--3 are plotted in the maximally general form as a function of the oscillation amplitude $x_0$ and applicable to the broad range of the field parameters $\omega_0$ and $E_0$. {Estimating the field parameters for a GaAs-based QPC with the Eckart potential barrier plotted in Fig.~1b, one can find, as an example, that the field (\ref{A}) with the frequency of THz range and the intensity of GW/cm$^2$ scale is appropriate to observe the discussed effects. Such an intensity can be achieved, particularly, for a pulsating field which was used in the recent experiments on the Floquet engineering of graphene~\cite{Cavalleri_2020}.}

It should be noted that any physically relevant repulsive potential driven by an oscillating field acquires the two-barrier structure if the field is both strong and high-frequency enough~\cite{Kibis_2019}. Therefore, the field-induced resonant tunneling through the quasi-stationary states confined between these two barriers is not unique feature of the Eckart potential (\ref{EP}) and expected to be in various quantum systems. Among them, the resonant tunnel decay of metastable quantum systems driven by an oscillating field should be noted especially. As an example, let us consider a particle confined between the two potential barriers pictured schematically in Fig.~4a. If the barriers are high and wide enough, the quasi-stationary state of the particle with energy $\varepsilon_0$ decays very slowly due to the weak tunneling through these barriers (the metastable state). In the presence of an oscillating field, each of these barriers acquires the two-barrier structure containing another quasi-stationary state with the energy $\varepsilon^\prime$ which can be tuned by the field (see Fig.~4b). When the resonant condition $\varepsilon_0=\varepsilon^\prime$ is satisfied, the metastable state decays very quickly due to the resonant tunneling (the resonant tunnel decay). Thus, an oscillating field can be considered as a tool to control the decay of various metastable quantum systems (e.g., the decay of electronic states confined in multi-barrier nanostructures).
\begin{figure}[h!]
\centering\includegraphics[width=0.7\columnwidth]{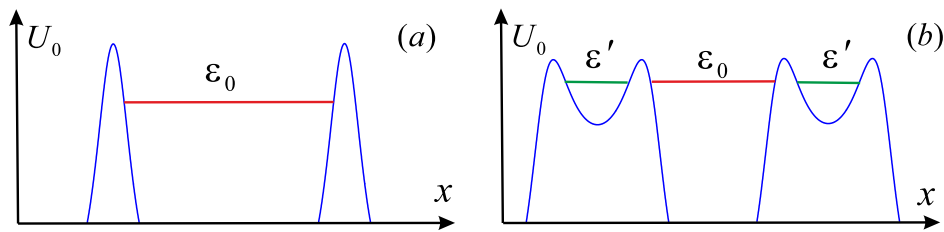}
\caption{Sketch of a metastable state with the energy $\varepsilon_0$ confined between the two potential barriers: (a) in the absence of an oscillating field; (b) in the regime of the resonant tunnel decay induced by the field.}
\end{figure}

\section*{Conclusion}
We developed the theory of the elastic electron tunneling through a potential barrier driven by a strong high-frequency electromagnetic field. The discussed effect of the field on the tunneling originates from the field-induced modification of the potential barrier. Particularly, the modified (dressed by the oscillating field) potential barrier acquires the two-barrier structure which contains quasi-stationary electron states confined between these two barriers. When the energy of an incident electron is equal to energies of these states, the resonant tunneling of the electron through the dressed barrier appears. The considered effect is of universal physical nature and can take place in the broad range of quantum systems. As possible manifestations of the effect, the resonant tunnel decay of metastable systems driven by an oscillating field and the features of electron transport through quantum point contacts irradiated by an electromagnetic wave should be noted.

\section*{Data availability}
All data generated or analysed during this study are included in this published article.

\section*{Acknowledgements}

We thank Z. D. Kvon for the fruitful discussion of quantum point contacts. The reported study was funded by the Russian Science Foundation (project 20-12-00001).

\section*{Author contributions}

O.V.K. formulated the physical problem under
consideration and wrote the main manuscript text, O.V.K. and V.M.K. derived analytical solutions of the problem, M.V.B. performed all numerical calculations and plotted figures. All authors taken part in discussions of used physical models and obtained results.

\section*{Additional information}

\textbf{Competing financial interests:} The authors declare no
competing financial interests.

\end{document}